\documentclass[lettersize,journal]{IEEEtran}
\usepackage{amsmath,amsfonts}
\usepackage{algorithmic}
\usepackage{algorithm}
\usepackage{array}
\usepackage[caption=false,font=normalsize,labelfont=sf,textfont=sf]{subfig}
\usepackage{textcomp}
\usepackage{stfloats}
\usepackage{url}
\usepackage{verbatim}
\usepackage{graphicx}
\usepackage{cite}
\begin{document}

\title{Homodyne vs. Heterodyne Architectures in Sub-THz Transceivers: A Phase Noise Perspective}

\author{\IEEEmembership{}
        \IEEEmembership{}        
\author{Didem Aydoğan,  Korkut Kaan Tokgöz\\
Faculty of Engineering and Natural Sciences, Electronics Engineering Department\\
Sabancı University, İstanbul, Türkiye \\
\small{ didem.aydogan@sabanciuniv.edu, korkut.tokgoz@sabanciuniv.edu }}
\thanks{This paper was produced by the IEEE Publication Technology Group. They are in Piscataway, NJ.}
\thanks{Manuscript received April 19, 2021; revised August 16, 2021.}}

\markboth{Journal of \LaTeX\ Class Files,~Vol.~14, No.~8, August~2021}%
{Shell \MakeLowercase{\textit{et al.}}: A Sample Article Using IEEEtran.cls for IEEE Journals}

\IEEEpubid{0000 0000/00\$00.00~\copyright~2021 IEEE}

\maketitle

\begin{abstract}
This letter examines the impact of oscillator phase noise on sub-terahertz OFDM
transceiver architectures, with a focus on the comparison between homodyne and
heterodyne designs. Using a Hexa-X compliant phase noise model, we analytically
show that heterodyne architectures reduce the total accumulated phase noise
variance by distributing frequency translation across lower-frequency
oscillators under realistic phase-noise scaling laws, thereby shifting the
dominant impairment from inter-carrier interference to common phase error. OFDM
simulations at 70 GHz and 140 GHz demonstrate that while homodyne architectures
remain competitive at mmWave frequencies, heterodyne designs provide improved
robustness to phase noise at higher sub-THz carriers. These results highlight
transceiver architecture as a key design lever for relaxing oscillator and
phase-locked loop constraints in future sub-THz wireless systems.
\end{abstract}


\section{Introduction}

Sub-terahertz (sub-THz) frequency bands have emerged as key enablers for
sixth-generation (6G) wireless systems, driven by the need for ultra-high data
rates, low latency, and integrated sensing and communication (ISAC)
capabilities \cite{HexaX}. Operating at carrier frequencies above 100 GHz, however,
introduces severe hardware challenges, among which oscillator phase noise (PN)
is one of the most critical\cite{Dband_PN,HW_IMP_modeling}. At these frequencies, high multiplication factors,
limited quality factors, and power-efficiency constraints lead to strong
phase noise that fundamentally limits system performance.

Orthogonal frequency-division multiplexing (OFDM), widely adopted due to its
spectral efficiency and flexibility, is particularly sensitive to phase noise.
The time-varying phase perturbation introduced by the local oscillator destroys
subcarrier orthogonality, giving rise to inter-carrier interference (ICI) and
common phase error (CPE). While numerous studies have focused on phase-noise
mitigation through waveform design or digital compensation\cite{eucnc_d,PN_COMPENSATION_WCL,WCL_FULLDUPpn}, the impact of the
transceiver architecture itself on phase-noise sensitivity has received
comparatively less attention, especially in the sub-THz regime.

\begin{figure}[b]
    \centering
    \includegraphics[width=1\linewidth]{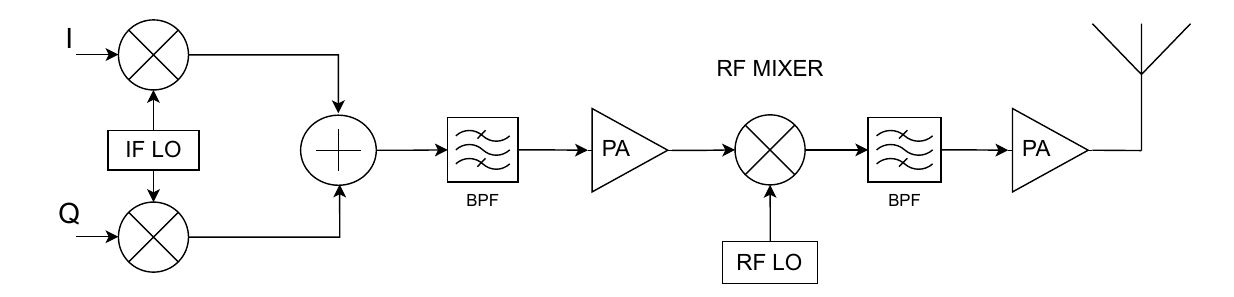}
    \caption{Heterodyne transmitter architecture.}
    \label{fig:heterodyne}
\end{figure}

In homodyne (direct-conversion) architectures, the local oscillator operates
directly at the radio-frequency (RF) carrier, often exceeding 140 GHz. Since
phase-noise variance scales quadratically with carrier frequency, such designs
suffer from strong ICI that is difficult to compensate digitally\cite{review-ARCHITECTURE}. In contrast,
heterodyne architectures as illustrated in Figure \ref{fig:heterodyne}  rely on one or more lower-frequency oscillators\cite{suberhetrodyne_THZ,heterodyne_60GHz},
splitting the frequency translation between an intermediate-frequency (IF)
stage and a final RF upconversion\cite{DSB_IF,wcl-lowif}. This architectural choice inherently reduces
the accumulated phase-noise variance and shifts the dominant impairment from
ICI to CPE, which can be efficiently mitigated using pilot-assisted digital
signal processing. Although direct-conversion architectures are widely used, future 6G and beyond communication and sensing technologies must reconsider these established approaches to cope with the growing severity of hardware impairments at higher carrier frequencies and to meet increasingly stringent system requirements.

This letter investigates the impact of transceiver architecture on phase-noise behavior in sub-THz OFDM systems, with a focus on homodyne versus heterodyne designs. Using realistic Hexa-X–compliant phase-noise models, we analytically characterize how the accumulated phase-noise variance depends on oscillator carrier frequency and intermediate-frequency (IF) placement, and how this dependence fundamentally differs across architectures. The analysis reveals architecture-dependent impairment regimes in which phase noise manifests predominantly as inter-carrier interference or common phase error. These insights are validated through OFDM simulations at 140 GHz, evaluating BER and EVM performance with and without pilot-based CPE correction. Overall, the results highlight transceiver architecture as a key design dimension for relaxing oscillator and phase-locked loop (PLL) constraints in sub-THz OFDM systems, without relying on complex waveform modifications or receiver side algorithms.
The remainder of this letter is organized as follows: Section II presents the phase-noise modeling framework and analyzes architecture-dependent phase-noise accumulation in homodyne and heterodyne OFDM transceivers. Section III discusses the architectural phase-noise impact and design guidelines, including IF placement. Section IV reports the simulation results and discussion, and Section V concludes the letter.
\section{Phase Noise Modeling and Architectural Impact}
\label{sec:pn_model}

\subsection{Zero Pole Phase Noise Modeling and Carrier-Frequency Scaling}
\label{subsec:zp_scaling}

In wideband sub-THz systems, simple white or Wiener phase noise models are
insufficient to capture realistic oscillator behavior. Instead, standardized
frameworks such as the Hexa-X and 3GPP phase noise model represent the oscillator spectrum
using cascaded zero-pole transfer functions \cite{HexaX},
\begin{equation}
S_{\varphi}(f_m) = S(0)\prod_{n=1}^{N_{stage}}
\frac{1+(f_m/f_{z,n})^2}{1+(f_m/f_{p,n})^2},
\label{eq:zp_model}
\end{equation}
where $S_{\varphi}(f_m)$ denotes the single-sideband phase-noise power spectral
density at offset frequency $f_m$ from the carrier, and $S(0)$ represents the
low-frequency phase-noise level. The product spans $N_{\mathrm{stage}}$
noise-shaping stages, where each pair $\{f_{z,n}, f_{p,n}\}$ corresponds to the
zero and pole corner frequencies associated with the PLL dynamics and oscillator
time constants. This generalized zero-pole representation, consistent with
Hexa-X and 3GPP phase-noise models, is adopted to ensure physically realistic
phase-noise behavior in sub-THz transceiver analysis.

A key property of oscillator phase noise is its dependence on the carrier
frequency. When scaling a reference phase noise model defined at
$f_{\text{base}}$ to a carrier frequency $f_c$, the PSD floor increases
quadratically as
\begin{equation}
S_{\varphi}(0)\big|_{f_c} =
\left(\frac{f_c}{f_{\text{base}}}\right)^2
S_{\varphi}(0)\big|_{f_{\text{base}}}.
\label{eq:fc_scaling}
\end{equation}
In addition to this amplitude scaling, the zero and pole corner frequencies
scale linearly with carrier frequency,
\begin{align}
f_{z,n}(f_c) &= f_{z,n}(f_{\text{base}})\cdot\frac{f_c}{f_{\text{base}}}, \\
f_{p,n}(f_c) &= f_{p,n}(f_{\text{base}})\cdot\frac{f_c}{f_{\text{base}}}.
\end{align}

This linear frequency shift is consistent with the behavior of frequency
multipliers of order $N_{multiplier}$, where the output phase noise can be approximated as
\begin{equation}
L_{\text{out}}(f_m) \approx L_{\text{in}}(f_m/N_{multiplier}) + 20\log_{10}(N_{multiplier}),
\end{equation}
indicating that frequency multiplication preserves the spectral shape while
increasing both the offset frequencies and the phase noise level. The entire PSD is shifted in frequency under
multiplication, a principle that underlies Hexa-X~\cite{HexaX}.

As a result, increasing the carrier frequency significantly raises the
integrated phase noise power, even when the zero pole structure remains
unchanged. This property directly motivates the use of lower-frequency
oscillators in heterodyne architectures, as will be quantified in the following
analysis.

\subsection{Phase Noise Accumulation in OFDM}

In OFDM systems, phase noise manifests as a time-varying phase rotation applied
to the transmitted samples. Over one OFDM symbol of duration
$T_{\text{sym}}=1/\Delta f$, the accumulated phase noise variance can be
approximated as
\begin{equation}
\sigma_{\varphi}^2
\approx \int_{0}^{B/2} S_{\varphi}(f_m)\,df_m,
\label{eq:pn_variance}
\end{equation}
where $B$ denotes the occupied bandwidth. This variance directly determines the
severity of phase noise distortion in the received signal.

At the receiver, the phase noise effect on subcarrier $k$ can be decomposed as
\begin{multline}
 Y_k =
X_k H_k
\underbrace{\frac{1}{N}\sum_{n=0}^{N-1} e^{j\varphi[n]}}_{\text{CPE}} \\
+ \sum_{\ell\neq k}
X_\ell H_\ell 
\underbrace{\frac{1}{N} 
\sum_{n=0}^{N-1} e^{j2\pi(\ell-k)n/N}e^{j\varphi[n]}}_{\text{ICI}} + N_k,
\end{multline}
where $Y_k$ denotes the received symbol on the $k$th subcarrier, $X_k$ and $H_k$
are the transmitted symbol and channel frequency response on subcarrier $k$,
respectively, and $\varphi[n]$ represents the discrete-time phase-noise process
at sample index $n$. The index $\ell$ denotes the interfering subcarrier index
($\ell \neq k$), $N$ is the number of subcarriers, and $N_k$ is additive noise.
The first term corresponds to the common phase error (CPE), which induces a
symbol-wide phase rotation, while the second term represents inter-carrier
interference (ICI) arising from phase-noise-induced loss of subcarrier
orthogonality.

ICI grows with the instantaneous phase fluctuation within the OFDM symbol and
is therefore highly sensitive to the total phase noise variance
$\sigma_{\varphi}^2$.
While both coherent phase error (CPE) and ICI scale with $\sigma_{\varphi}^2$, ICI is
particularly sensitive to high-frequency phase fluctuations. By reducing the
total phase noise variance through heterodyne operation, the dominant impairment
shifts from ICI to CPE, which can be efficiently compensated using low-complexity
pilot-based phase estimation.

\subsection{Homodyne versus Heterodyne Architectures}\label{subsec:hom_vs_het}

In homodyne (direct-conversion) architectures, the local oscillator operates
directly at the RF carrier frequency $f_{\text{RF}}$, such that the effective
phase noise PSD equals that of the RF oscillator,
\begin{equation}
S_{\varphi}^{\text{hom}}(f_m) = S_{\varphi,\text{LO}}(f_m)\big|_{f_{\text{RF}}}.
\end{equation}
Due to the quadratic scaling in \eqref{eq:fc_scaling}, this results in large
$\sigma_{\varphi}^2$ at sub-THz frequencies, leading to dominant ICI that is
difficult to mitigate digitally.

In contrast, heterodyne architectures split the frequency translation between
an intermediate-frequency (IF) oscillator at $f_{\text{IF}}$ and a second RF
local oscillator at $f_{\text{RF}}-f_{\text{IF}}$. Assuming independent phase
noise processes, the total phase noise variance becomes
\begin{equation}
\sigma_{\varphi,\text{het}}^2
=
\sigma_{\varphi}^2(f_{\text{IF}})
+
\sigma_{\varphi}^2(f_{\text{RF}}-f_{\text{IF}}),
\label{eq:het_variance}
\end{equation}
where each term follows the scaling law in \eqref{eq:fc_scaling}. This structure
naturally reduces the accumulated phase noise compared to homodyne operation,
since both oscillators operate at lower frequencies than $f_{\text{RF}}$.

Equation~\eqref{eq:het_variance} further reveals that the total phase noise
variance exhibits a convex dependence on $f_{\text{IF}}$, with a minimum when
the IF and RF-LO frequencies are symmetrically placed around $f_{\text{RF}}/2$.
Beyond reducing $\sigma_{\varphi}^2$, this architectural choice shifts the
dominant impairment from ICI toward CPE, which can be efficiently compensated
using low-complexity, pilot-based phase estimation.

These analytical insights form the basis for the simulation results presented
in the next section, where Hexa-X-compliant phase noise models are used to
quantify the performance difference between homodyne and heterodyne OFDM
transceivers at sub-THz frequencies.

\section{Architectural Phase Noise Impact and Design Guidelines}
\label{subsec:arch_pn_guidelines}

\subsection{Homodyne versus Heterodyne Phase Noise Accumulation}
As discussed in Section \ref{subsec:hom_vs_het}, homodyne (direct-conversion) architectures employ a
local oscillator operating directly at the RF carrier frequency
$f_{\mathrm{RF}}$. Due to the quadratic carrier-frequency scaling of oscillator
phase noise, this results in a large accumulated phase-noise variance
$\sigma_{\varphi}^2$ and consequently in ICI-dominated distortion that is
difficult to mitigate digitally. In contrast, heterodyne architectures
distribute the frequency translation between an intermediate-frequency (IF)
oscillator at $f_{\mathrm{IF}}$ and a second RF local oscillator at
$f_{\mathrm{RF}}-f_{\mathrm{IF}}$.

Assuming independent phase noise
processes, the total phase noise variance accumulated over one OFDM symbol can be
expressed as
\begin{equation}
\sigma_{\varphi}^2
=
\sigma_{\varphi,\mathrm{IF}}^2
+
\sigma_{\varphi,\mathrm{RF\text{-}LO}}^2,
\label{eq:sigma_split}
\end{equation}
where each term follows the Hexa-X carrier-frequency scaling law introduced in
Section~II.

Using the linear PSD scaling from a reference frequency $f_{\text{base}}$, the
variance contribution of an oscillator operating at frequency $f_c$ is
approximated as
\begin{equation}
\sigma_{\varphi}^2(f_c)
\approx
N_{sample}\,2\pi\Delta f\,S_0
\left(\frac{f_c}{f_{\text{base}}}\right)^2,
\end{equation}
where $N_{sample}$ is the number of samples per OFDM symbol and $S_0$ denotes the
reference PSD level.

\subsection{Optimal IF Placement and Phase Noise Reduction}

\begin{figure}[h!]
    \centering
    \includegraphics[width=0.9\linewidth]{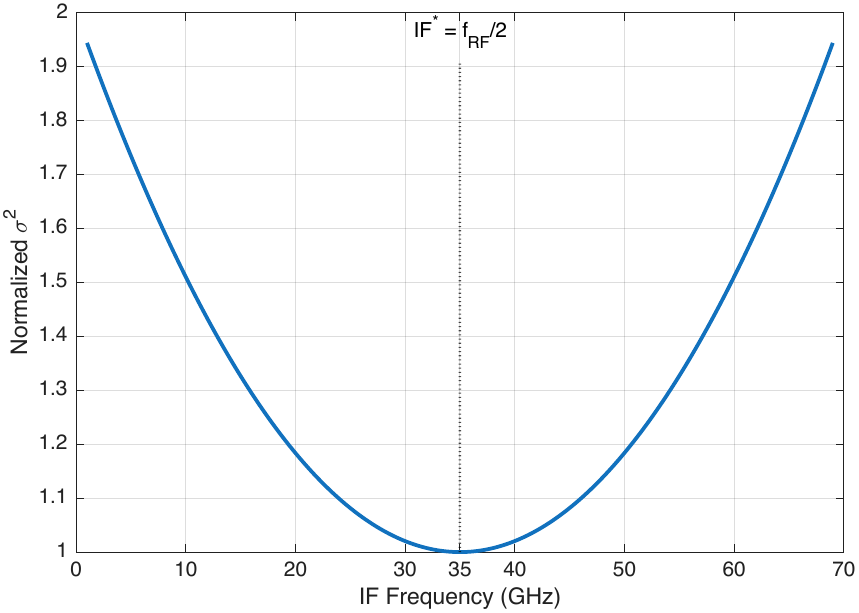}
    \caption{Total phase noise variance vs. IF frequency for target RF of 70 GHz.}
    \label{fig:sigma70}
\end{figure}

\begin{figure}[h!]
    \centering
    \includegraphics[width=0.9\linewidth]{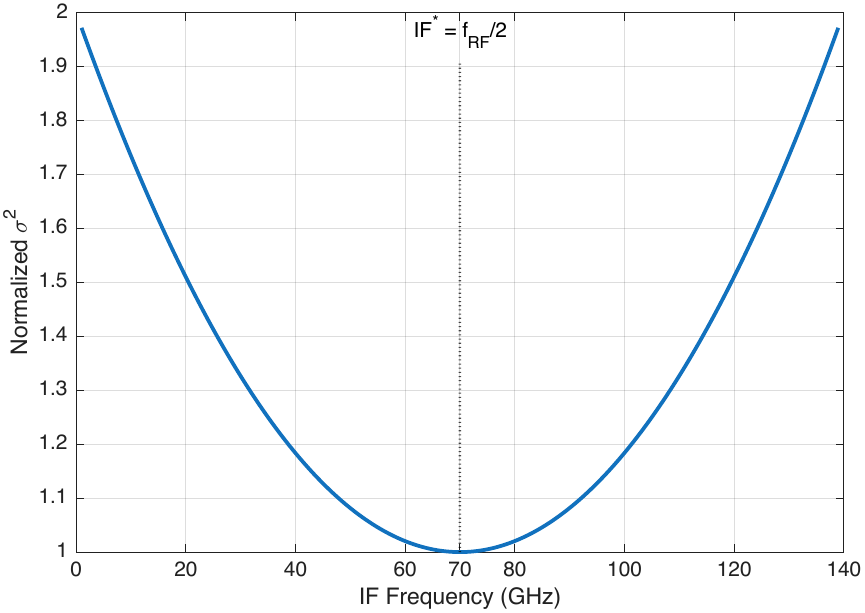}
    \caption{Total phase noise variance vs. IF frequency for target RF of 140 GHz (right).}
    \label{fig:sigma140}
\end{figure}

For a fixed target RF carrier $f_{\mathrm{RF}}$, the IF and RF-LO frequencies
satisfy
\begin{equation}
f_{\mathrm{RF}} = f_{\mathrm{IF}} + f_{\mathrm{RF\text{-}LO}}.
\end{equation}
Substituting into \eqref{eq:sigma_split} yields
\begin{equation}
\sigma_{\varphi}^2
=
N_{sample}\,2\pi\Delta f\,S_0
\!\left[
\left(\frac{f_{\mathrm{IF}}}{f_{\text{base}}}\right)^2
+
\left(\frac{f_{\mathrm{RF}}-f_{\mathrm{IF}}}{f_{\text{base}}}\right)^2
\right],
\label{eq:sigma_if}
\end{equation}
which exhibits a convex dependence on $f_{\mathrm{IF}}$. The total phase noise
variance is minimized when the IF and RF-LO frequencies are symmetrically placed
around $f_{\mathrm{RF}}/2$. Figure \ref{fig:sigma70} and Figure \ref{fig:sigma140} illustrate the analytically predicted total phase
noise variance $\sigma_{\varphi}^2$ as a function of the intermediate frequency (IF)
for two target RF carriers, 70 GHz and 140 GHz. In both cases, a clear U-shaped
dependence is observed, confirming that the total variance is the sum of the
independent contributions of the IF and RF local oscillators. The minimum variance occurs when the IF and RF-LO frequencies are symmetrically
placed around the target RF carrier, i.e., near 35 GHz for a 70 GHz carrier and
near 70 GHz for a 140 GHz carrier. This behavior directly validates the proposed
heterodyne design guideline: splitting the frequency translation across two
lower-frequency oscillators minimizes the accumulated phase noise variance
compared to homodyne operation, where a single oscillator operates at the full
RF carrier. These results provide a clear design guideline: distributing the frequency
translation across two lower-frequency oscillators significantly reduces the
accumulated phase-noise variance compared to homodyne operation. 
Specifically,
a heterodyne architecture employing an IF oscillator at $f_{\mathrm{IF}}$
achieves a variance reduction factor
\begin{equation}
\gamma
=
\frac{\sigma_{\varphi,\mathrm{IF}}^2}{\sigma_{\varphi,\mathrm{hom}}^2}
=
\left(\frac{f_{\mathrm{IF}}}{f_{\mathrm{RF}}}\right)^2,
\end{equation}
corresponding to an $82\%$ reduction for $f_{\mathrm{IF}}=30$ GHz and
$f_{\mathrm{RF}}=70$ GHz.

This scaling has direct implications for sub-THz VCO and PLL design. While
homodyne architectures rely on a single high-frequency oscillator often
requiring large multiplication factors and incurring a $20\log_{10}(N)$ phase
noise penalty, heterodyne systems enable the use of lower-frequency IF
oscillators with inherently improved phase noise and relaxed loop design
constraints. Consequently, heterodyne architectures offer a practical and
energy-efficient approach to mitigating oscillator phase-noise limitations in
sub-THz OFDM systems.

\section{Results and Discussion}
\label{subsec:sim_framework}

To complement the analytical insights on $\sigma_{\varphi}^2$, we evaluate the
communication impact of phase noise using bit error rate (BER) and error vector
magnitude (EVM). A complete OFDM transceiver is simulated under Hexa-X phase
noise, comparing homodyne and heterodyne frequency plans with and without
pilot-aided common phase error (CPE) correction.

\subsubsection{Simulation Parameters}

All simulations use the same OFDM numerology and Monte Carlo configuration,
summarized in Table~\ref{tab:sim_params}. The carrier frequency is set to
$f_c\in\{70,140\}$ GHz for homodyne operation. For heterodyne operation, the RF
carrier is generated by combining two equal-frequency oscillators:
$35{+}35$ GHz for the 70 GHz case and $70{+}70$ GHz for the 140 GHz case. Each
scenario is simulated both \emph{without} CPE correction and \emph{with}
pilot-based CPE correction.

\begin{table}[b]
\caption{Simulation configuration for BER/EVM evaluation.}
\label{tab:sim_params}
\centering
\begin{tabular}{|l|c|}
\hline
\textbf{Parameter} & \textbf{Value} \\
\hline
Architectures / carriers &
\begin{tabular}[c]{@{}l@{}}
Homodyne: $f_c=70$ GHz, $140$ GHz \\
Heterodyne: $35{+}35$ GHz $\rightarrow$ 70 GHz, \\
\hspace{15mm}$70{+}70$ GHz $\rightarrow$ 140 GHz
\end{tabular}
\\
\hline
Subcarriers ($N$) & $256$ \\
\hline
Cyclic prefix length & $N/16$ samples \\
\hline
OFDM symbols per run & $1000$ \\
\hline
Modulation & 64-QAM ($k=\log_2 64$) \\
\hline
Subcarrier spacing & $\Delta f = 240$~kHz \\
\hline
SNR sweep & 0:1:30 dB \\
\hline
Monte Carlo runs & $N_{\text{mc}}=200$ \\
\hline
Pilot fraction for CPE & $1/4$ (uniformly spaced) \\
\hline
Phase noise model & Hexa-X, scaled from 15 GHz reference \\
\hline
\end{tabular}
\end{table}

\subsubsection{Phase Noise Injection and OFDM Processing}

Let $s[n]$ denote the discrete-time OFDM sample sequence at the transmitter.
Phase noise is applied sample-wise as
\begin{equation}
s_{\text{pn}}[n] = s[n]e^{j\phi[n]},
\end{equation}
where $\phi[n]$ is generated according to the adopted Hexa-X phase noise model
(scaled to the corresponding oscillator frequency). At the receiver, the cyclic
prefix is removed and FFT demodulation is applied; phase noise then manifests as
a combination of CPE and ICI, degrading both detection and constellation quality.

\subsubsection{Pilot-Aided CPE Correction}

When enabled, CPE is estimated from pilot subcarriers as
\begin{equation}
\hat{\phi}_{\text{CPE}} =
\angle\!\left(\frac{1}{N_p}\sum_{k\in\mathcal{P}} Y_k X_k^{*}\right),
\end{equation}
where $\mathcal{P}$ is the pilot set, $N_p=|\mathcal{P}|$, $X_k$ are known pilot
symbols, and $Y_k$ are the received subcarriers. The received OFDM symbol is then
compensated by
\begin{equation}
Y_k^{\text{comp}} = Y_k e^{-j\hat{\phi}_{\text{CPE}}}, \quad \forall k.
\end{equation}

\begin{figure}[b] \centering \includegraphics[width=0.9\linewidth]{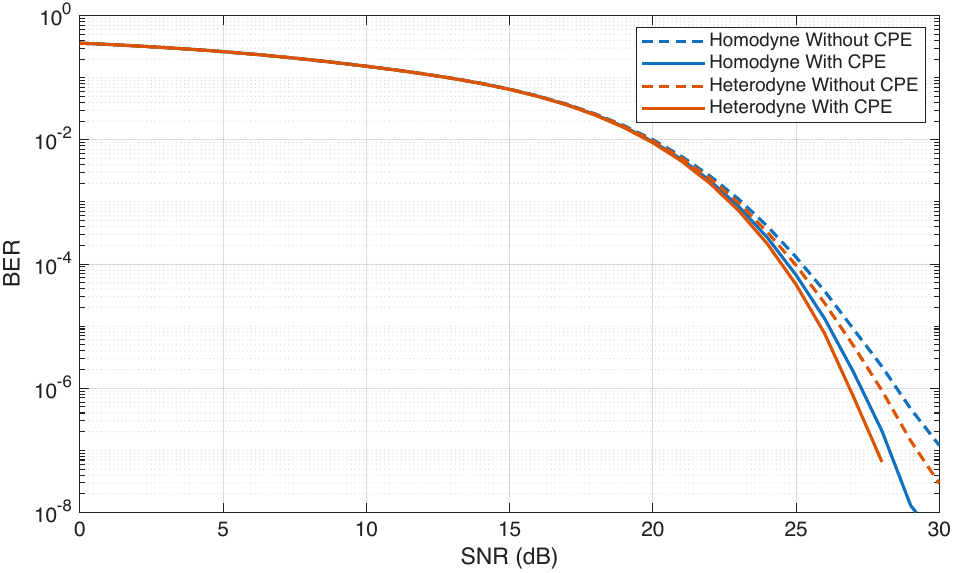} \caption{BER at 70 GHz carrier frequency comparing heterodyne and homodyne systems under different CPE compensation strategies.} \label{fig:ber_70} \end{figure}

\begin{figure}[t] \centering \includegraphics[width=0.9\linewidth]{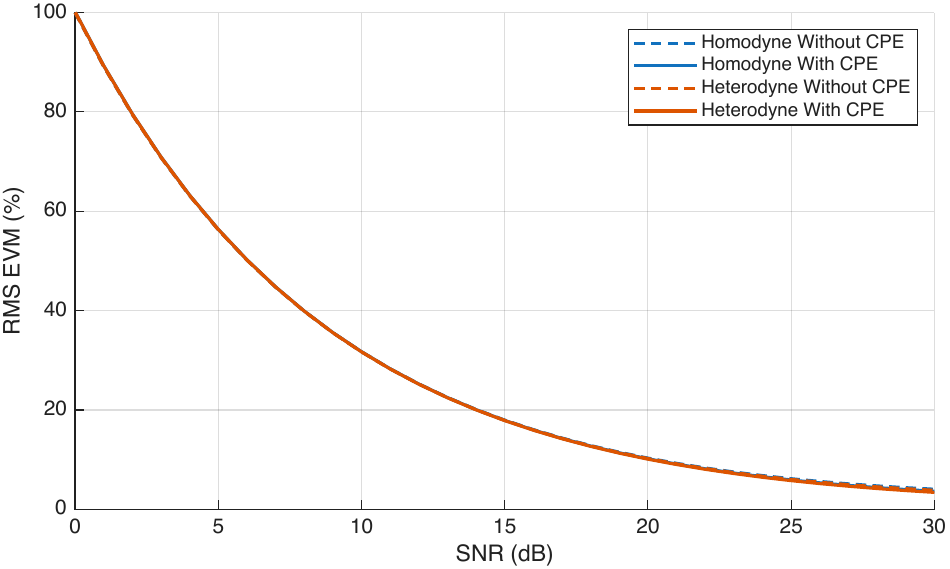} \caption{EVM at 70 GHz carrier frequency comparing heterodyne and homodyne systems under different CPE compensation strategies.} \label{fig:evm_70} \end{figure}

Fig.~\ref{fig:ber_70} and Fig.~\ref{fig:evm_70} report the BER and EVM performance
at a 70 GHz carrier frequency. The homodyne architecture is compared against a
heterodyne configuration using two 35 GHz oscillators. Results are shown both
without and with pilot-based common phase error (CPE) compensation.

\begin{figure}[t] \centering \includegraphics[width=0.9\linewidth]{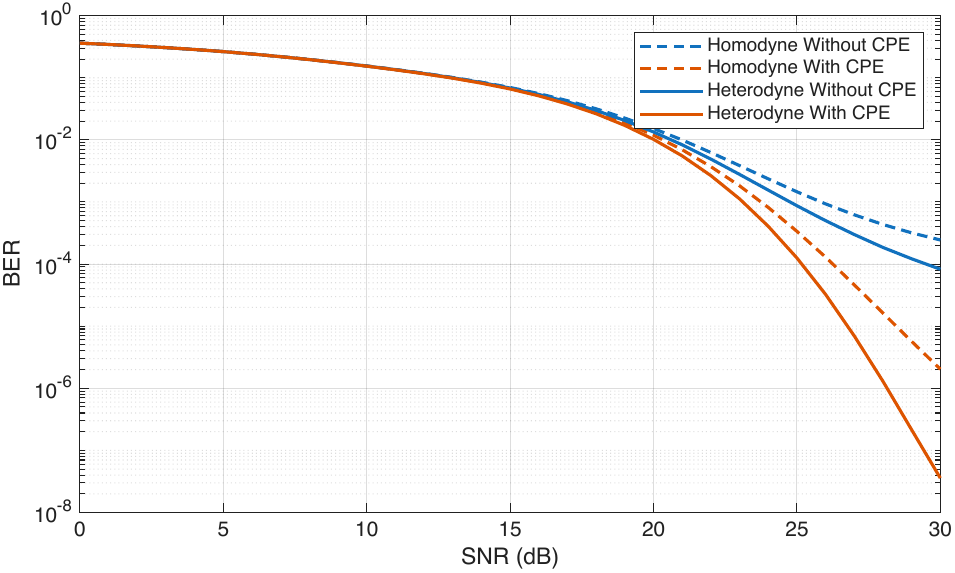} \caption{BER at 140 GHz carrier frequency comparing heterodyne and homodyne systems under different CPE compensation strategies.} \label{fig:BER_140} \end{figure}

\begin{figure}[h!] \centering \includegraphics[width=0.9\linewidth]{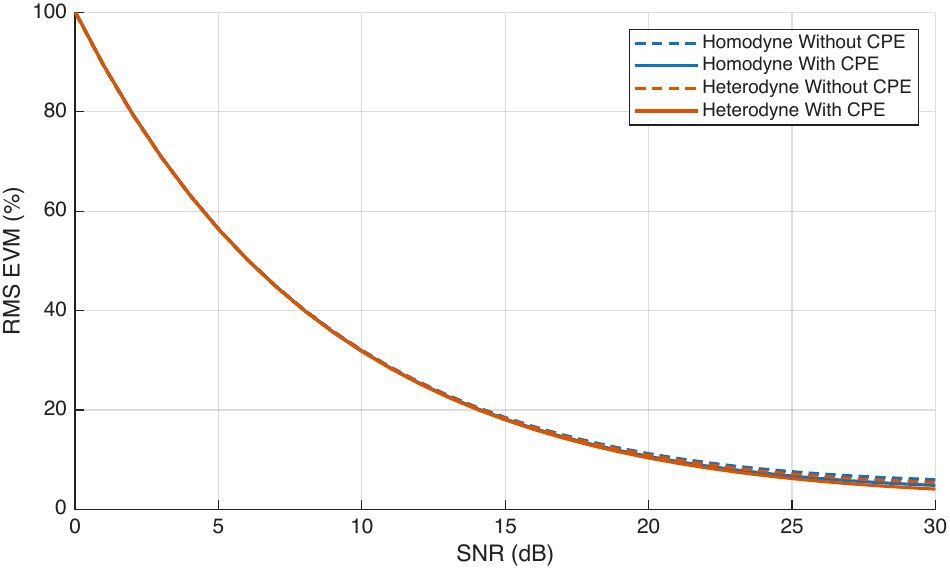} \caption{EVM at 140 GHz carrier frequency comparing heterodyne and homodyne systems under different CPE compensation strategies.} \label{fig:evm_140} \end{figure}

Without CPE correction, the heterodyne system already outperforms the homodyne
case in both BER and EVM, reflecting its lower inherent phase noise variance.
When CPE correction is enabled, both architectures benefit from improved
performance; however, the heterodyne configuration achieves a steeper BER decay
and lower residual EVM across the entire SNR range. At high SNR values, the
heterodyne system with CPE correction reaches significantly lower BER than the
homodyne counterpart, indicating that reduced inter-carrier interference (ICI)
is the dominant factor limiting performance at this frequency.

The impact of architecture choice becomes more pronounced at higher carrier
frequencies. Fig.~\ref{fig:BER_140} and Fig.~\ref{fig:evm_140} present the BER and
EVM results at 140 GHz, comparing a homodyne transceiver against a heterodyne
implementation using two 70 GHz oscillators. At 140 GHz, the homodyne system experiences severe degradation due to the
quadratic increase of phase noise with carrier frequency. This manifests as
higher BER and larger EVM, particularly at moderate-to-high SNR values. In
contrast, the heterodyne architecture consistently achieves lower BER and EVM,
both before and after CPE correction, owing to its reduced total phase noise
variance. CPE compensation remains effective for both architectures; however, its benefit
is more pronounced in the heterodyne case, where phase noise distortion is
primarily CPE-dominated rather than ICI-dominated. These results confirm that,
at sub-THz frequencies, heterodyne architectures not only reduce the overall
phase noise impact but also shift the impairment structure toward forms that are
more amenable to low-complexity digital compensation.\\

Across both carrier frequencies, the results consistently demonstrate that
heterodyne architectures provide improved robustness to phase noise compared to
homodyne designs. The gains observed in BER and EVM directly follow the reduction
in total phase noise variance predicted by the analytical model and illustrated
by the IF-frequency sweep.

From a system design perspective, these results highlight that architectural
choices can significantly relax oscillator phase noise requirements without
relying on complex signal processing. By operating oscillators at lower
frequencies and balancing the IF and RF-LO contributions, heterodyne
architectures inherently suppress ICI and enable efficient pilot-based CPE
correction. This makes heterodyne design a practical and scalable solution for
OFDM-based communication at mmWave and sub-THz frequencies.

\section{Conclusion}
This letter investigated the impact of transceiver architecture on phase
noise sensitivity in sub-THz OFDM systems. Using a Hexa-X compliant phase noise
model, we showed that heterodyne architectures reduce the total accumulated phase
noise variance by distributing frequency translation across lower-frequency
oscillators, thereby shifting the dominant impairment from inter-carrier
interference to common phase error. Simulation results at 70 GHz and 140 GHz confirm that the benefit of heterodyne
operation is strongly frequency-dependent. While homodyne architectures remain
competitive at mmWave frequencies, heterodyne designs provide superior
robustness to phase noise at higher sub-THz carriers, enabling more effective
CPE compensation. These results highlight heterodyne architecture as a practical
means to relax oscillator and PLL requirements for future sub-THz OFDM-based 6G
systems.

\bibliographystyle{IEEEtran}
\bibliography{bibliography}

\vfill

\end{document}